\documentstyle[psfig,aps,preprint]{revtex}

\begin{document}
%\draft

%%% Turns off the header ``FIGURES'' above the first figure. %%%
%%% Use for preprint version.                                %%%
\global\firstfigfalse

\title{Transfer and Decay of an Exciton Coupled to Vibrations in a Dimer}

\author{Holger Schanz} 
\address{Institut f\"ur Physik, Humboldt-Universit\"at, 
Invalidenstr. 110\\ 10~115 Berlin, Germany\\
e-mail: schanz@physik.hu-berlin.de} 

\author{Ivan Barv\'{\i}k} 
\address{Institute of Physics, Charles University, 
Ke Karlovu 5\\ 121~16 Prague, Czech Republic} 

\author{Bernd Esser} 
\address{Institut f\"ur Physik, Humboldt-Universit\"at, 
Invalidenstr. 110\\ 10~115 Berlin, Germany} 

\date{June 14, 1996}

\maketitle
\thispagestyle{empty}

\begin{abstract}
Transfer and decay dynamics of an exciton coupled to a polarization vibration
in a dimer is investigated in a mixed quantum-classical picture with
the exciton decay incorporated by a sink site.  Using a separation of
time scales, it is possible to explain analytically the most important
characteristics of the model. 

If the vibronic subsystem is fast, these are the enhancement of
nonlinear self trapping due to the sink and the slowing down of the
exciton decay for large coupling or sink strength.  Numerical results
obtained recently for the DST approximation to the model are
quantitatively explained and new dynamic effects beyond this
approximation are found. 

If the vibronic subsystem is slow, the behavior of the system follows
closely the predictions of the adiabatic approximation.  In this
regime, the exciton decay crucially depends on the initial conditions
of the vibronic subsystem.  

In the transition regime between adiabatic and DST approximation,
complex dynamics is observed by numerical computation. We discuss the
correspondence to the chaotic behavior of the excitonic-vibronic
coupled dimer without trap.
\end{abstract}

\thispagestyle{empty}

\newcommand{\nc}{\newcommand}
\newcommand{\rnc}{\renewcommand}

\nc{\mrm}{\rm}
\nc{\mbf}{\bf}

\nc{\sep}{\hspace*{1cm}}
\nc{\be}{\begin{equation}}
\nc{\bel}{\begin{equation}\label}
\nc{\ee}{\end{equation}}
\nc{\bea}{\begin{eqnarray}}
\nc{\eea}{\end{eqnarray}}

\nc{\eps}{\epsilon}
\nc{\veps}{\varepsilon}
\nc{\ga}{\gamma}
\nc{\Ga}{\Gamma}
\nc{\om}{\omega}

\rnc{\/}{\over}
\nc{\half}{{1\/2}}
\rnc{\(}{\left (}
\rnc{\)}{\right )}
\rnc{\[}{\left [}
\rnc{\]}{\right ]}

\nc{\rf}{fig.~\ref}

\section{Introduction}
The purpose of this paper is to study the interplay between a coherent
transfer regime of an exciton and two processes leading to the loss of
the linear character of the exciton transfer, namely trapping of the
exciton at a sink site with a prescribed sink rate $ \Gamma$ and the
coupling to intramolecular polarization vibrations.  

A lot of work has been done on exciton transfer theories during the
last decades.  Beginning with the microscopic treatment by Haken and
Reineker \cite{HR68} and Grover and Silbey \cite{GS70} a number of
theories such as the Continuous Time Random Walk (CTRW) \cite{KMS73},
the Pauli Master equation (PME) \cite{KR75}, the Generalized Master
equation (GME) \cite{Ken82}, the Stochastic Liouville equation (SLE)
and the Haken-Strobl-Reineker (HSR) model (see \cite{Rei82} and
references therein) were developed and mainly directed to obtaining
equations which describe the coupled coherent and incoherent motion of
the excitation. The investigation of exciton transfer on relatively
small molecular aggregates such as dimers or triades is of much
interest for clarifying the applicability of exciton transfer theories
to experimental situations such as described in \cite{OMY90,RMB90}.

Trapping of quasiparticles due to a sink site constitutes an important
phenomenon in many molecular systems. In photosynthesis, for instance,
an exciton in a harvesting antenna transfers its energy to a reaction
center, where it can be trapped. Electron transfer processes then
follow.  Pearlstein and his coworkers were the first who recognized
that the consequences of a sink on the energy transfer processes are
different in the coherent and incoherent regimes \cite{PZ85} (and
references therein).  \v{C}\'{a}pek and Sz\"{o}cs \cite{CS84} pointed
out the necessity of a transformation of the memory functions in the
presence of a sink. They also gave a prescription for a proper
inclusion of the sink into the HSR model. This found application
e.~g.\ in computer simulations of the excitation transfer in
photosynthetic systems \cite{Bar91,Bar91a}.  Recently, a new form of
purely coherent memory functions in presence of a sink was derived and
shown to have important  consequences for the excitation transfer
\cite{BH}.  New memory functions were also used to obtain
characteristics entering the CTRW description in the presence of a
sink \cite{HB}.

The coupling between electronic and vibronic degrees of freedom in
molecular and condensed media is another basic mechanism influencing
transfer properties of electronic excitations in these systems. The
investigation of its consequences started from the polaron problem in
solid states (see e. g. \cite{App68} and references therein; for
exciton-phonon interaction see \cite{Dav79}) and continued with the
study of the influence of the vibronic bath variables on the
excitation transfer properties in the framework of the generalized
Master equation \cite{Ken82} and stochastic Liouville equation
approaches \cite{Rei82}. 

With the development of the theory of dynamical systems it has become
attractive to analyze the implications of electronic-vibronic
couplings employing concepts and methods of this field. Using such a
dynamic system approach we study the detailed picture of the time
evolution of a small number of relevant variables of the system, which
are assumed to interact weakly with the environment. Recent
experimental developments in the field of ultrashort time resolved
spectroscopy (see e.\ g.\ \cite{Q93}) seem to make a direct
observation of this time evolution possible in the near future.

In the present paper we focus our interest on the exciton dynamics in
a molecular dimer coupled to intramolecular vibrations.  A remarkable
feature of this model is the possibility of self trapping, i.~e.
unequal time averaged occupation probabilities on the two sites of the
configuration.  The easiest way to obtain this effect from a coupling
to vibrational degrees of freedom leads to the two-site discrete self
trapping (DST) equation \cite{ELS85,KC86,EH91} which is a nonlinear
but self contained equation of motion for the excitonic site
occupation amplitudes. More sophisticated approaches take the dynamics
of the vibrations explicitly into account by using a mixed
quantum-classical description \cite{KK94a,ES} or by treating the
coupled system quantum mechanically \cite{SE96}. Another interesting
point is the effect of dissipation on self trapping \cite{KWu89,VAG94,SBR95}.

Although the influences of vibronic coupling and trapping on transport
properties have been investigated separately in great detail, the
combination of both, which can be important in the application of
transfer theory, has rarely been addressed in the past.  In a recent
paper \cite{BES95} we have studied the interplay between vibrational
coupling and trapping due to a sink site for a dimer and a trimer in the
framework of the DST approximation.  An extension of the results seems
to be possible into two directions.  One can either try to describe
larger molecular aggregates within the same framework or one can try to
improve the dynamic model which is used. In this paper we will adopt the
latter way and take account of the dynamics of a particular vibronic
variable, too.  For this purpose the coupled excitonic-vibronic system
will be treated in a mixed quantum-classical description
\cite{KK94a,ES}, which is justified whenever the quantum fluctuations in
the vibronic subsystem are negligible.
The model we use will be specified and developed further in
section \ref{2}, where we also indicate the modifications that lead to the DST
equation. 

The explicit introduction of a vibronic degree of freedom makes not only
the dynamics of the system more complex, it also increases the number of
initial conditions that have to be specified in order to uniquely
determine a solution. Motivated by the experimental situation e.~g.\ in
photosynthetic units, the creation of the exciton will always be assumed 
at the site without sink. For the vibronic degrees of freedom we will
consider various possibilities.  As we shall see, the initial
conditions can crucially influence the dynamics and it is not easy to
overlook all the dynamical regimes the model is capable of from
numerical computations only.  Therefore we devote section \ref{3} to an
analytical treatment of the system assuming some separation between the
different time scales.  In particular, we will generalize the analysis
of the fixed points for the excitonic-vibronic coupled dimer without
sink \cite{ES} to the present system. This will provide a global picture
of the phase space which we shall further support in section \ref{4} by
discussing the time dependence of the total occupation probability and
the relative site occupation probabilities for some specific solutions. 

The results of section \ref{3} will also allow for a quantitative
understanding of the DST dimer with sink.  One of our aims is to study
what is inherited in our more complex model from this approximation and
which dynamic effects are beyond it. We will therefore briefly recall
the relevant numerical results from \cite{BES95} and discuss them in the
light of our findings in section \ref{3.1}.  

Since the mixed quantum-classical description we are using in this paper
can be justified best for small oscillator frequencies, we pay much
attention to this adiabatic regime, too. As we shall see, the assumption
of slow vibrations is just the antipode of the DST case and therefore
new dynamic effects different from DST results can be expected from it. 

%%%%%%%%%%%%%%%%%%%%%%%%%%%%%%%%%%%%%%%%%%%%%%%%%%%%%%%%%%%%%%%%%%%%%%

\section{Description of the Model}
\label{2}
\subsection{The Hamiltonian}
We consider the dynamics of an exciton moving on a molecular dimer.
At each of the two monomers the exciton is allowed to interact with
an intramolecular vibronic degree of freedom. Thus, the Hamiltonian of
our model contains excitonic, vibronic and interaction parts denoted
by $H_{\mrm exc}$, $H_{\mrm vib}$ and $H_{\mrm int}$, respectively:
\begin{equation}
\label{M1}
H = H_{\mrm exc} + H_{\mrm vib} + H_{\mrm int},
\end{equation}
$H_{\mrm exc}$ describes a two site model
\begin{equation}
\label{M2}
H_{\mrm exc} = \sum_{n} \epsilon_n c_n^* c_n + \sum_{n\ne m} V_{nm} c_n^* c_m
\end{equation}
with $n, m \;=\; 1,2$. $c_n$ is the probability amplitude of the
exciton to occupy the n-th molecule and $V_{nm}$ the transfer matrix
element due to dipole-dipole interaction. In a standard two site model,
the $\eps_i$ are real quantities and correspond to the local site energies of
the exciton. Here, we allow $\eps_2$ to contain a negative imaginary
part in order to describe the decay of the exciton on the sink site 2.
Since we are not interested in the effect of a site energy difference
on the exciton dynamics here, we set
\be
\eps_1 = 0\sep \eps_2 = - i{\Ga\/2},
\ee
which is equivalent to the extended sink model for the exciton decay
introduced in \cite{CS84} on the density matrix level. This model has
been shown to solve the problem of a consistent description of 
exciton trapping at a sink meeting basic physical requirements such
as positive occupation probabilities.

The vibrational part $H_{\mrm vib}$ is taken as the sum of the energies
corresponding to intramolecular vibrations at each of the monomers for
which we use the harmonic approximation
\begin{equation}
\label{M3}
H_{\mrm vib} = \sum_n \frac{1}{2}(p_n^2 + \omega_n^2 q_n^2).
\end{equation}
Here $q_n ,p_n$ and $\omega_n$ are the coordinate, the canonic conjugate
momentum and the frequency of the intramolecular vibration of the $n$-th
molecule, respectively.

The interaction Hamiltonian takes into account that the exciton energy
depends on the molecular configuration of the monomers which is
expressed by the coordinates $q_n$. Using a first order expansion in
$q_n$ one has
\begin{equation}
H_{\mrm int} = \sum_n \gamma_n q_n c_n^* c_n,
\label{M4}
\end{equation}
where $\gamma_n$ are some coupling constants.  In order to restrict the
number of free parameters as much as possible we will assume that the
dimer is symmetric except for the additional sink term on site 2,
i.~e.~$\om_1 = \om_2 = \om$, $\ga_1 = \ga_2 = \ga$ and $V_{12}=V_{21}=-V$.

In what follows we will use a mixed quantum-classical description of
the dynamics, i.~e.~we treat the vibronic degrees of
freedom in the classical approximation while retaining the quantum wave
function for the excitonic two site system.  This approximation can be
justified over a finite time range which increases as the oscillator
frequencies and coupling constants decrease. When it exceeds the
life time of the exciton the mixed quantum-classical picture describes
correctly the decay of the excitation.

Using units with $\hbar = 1$ we obtain from (\ref{M1})-(\ref{M4})
the equations of motion
\begin{eqnarray}
\label{eom_c1} i \,d/dt\,{c}_1 &=& \gamma q_1 c_1 - V\,c_2 \\
\label{eom_c2} i \,d/dt\,{c}_2 &=& \(-i{\Ga\/2} + \gamma q_2\) c_2 - V\,c_1 \\
\label{eom_q} \,d/dt\,{q}_n &=& p_n \\
\label{eom_p} \,d/dt\,{p}_n &=& - \omega^2 q_n - \gamma |c_n|^2.
\end{eqnarray}
% In the following we will use the parameter 
% \be
% \chi := \({\ga\/\om}\)^2
% \ee
% instead of $\ga$ to describe the coupling strength, since this
% parameter has been shown to control the qualitative behavior of the
% solutions when no sink is present \cite{ES} and this choice will
% facilitate the comparison of our results to the DST approximation in
% \cite{BES95}.

\subsection{Reduced Equations of Motion}
For a numerical investigation the equations (\ref{eom_c1})-(\ref{eom_p}) are well
suited and we have integrated them in order to obtain the results that
will be presented in section \ref{4}.  The analytical treatment of section \ref{3},
however, requires to reduce the number of variables and free parameters
as much as possible. Therefore we will now 
rewrite the equations of motion (\ref{eom_c1})-(\ref{eom_p}) using
appropriate dimensionless variables and parameters.

The excitonic subsystem can be described by a point on the Bloch
sphere which is usually given in Cartesian coordinates. Here we
prefer to parametrize the Bloch sphere in spherical coordinates $R$,
$\theta$ and $\phi$. They are defined using the density matrix of the
two-site system $\rho_{mn} = c_m c_n^*$ ($n,m=1,2$) 
\begin{eqnarray}
R \, := & \rho_{11} + \rho_{22}& \nonumber \\
\cos\theta\;\,R\, := & \rho_{22} - \rho_{11} &\hspace*{5mm} (0
\le \theta \le \pi) \\
e^{i\phi}\sin\theta \;\,R\,   := & 2 \rho_{12} &\hspace*{5mm} (-\pi <
\phi \le \pi)\nonumber \;.
\end{eqnarray}
Due to the trapping of the exciton the radius of the Bloch sphere
$R(t)$ which is the total probability to find an exciton on either of
the two sites is not constant but a monotonically decreasing
function of time with $R(0) = 1$. 

Besides the total occupation probability $R$ the difference of the
occupation of the two sites is of interest.
It is determined by the angle $\theta$ since we have
\be\label{sod}
|c_1|^2 = {1-\cos\theta\/2}R \sep |c_2|^2 = {1+\cos\theta\/2}R \,. 
\ee
The phase $\phi$ has no direct physical interpretation. We note that
$\phi$ is not well defined at the points $\theta = 0$ and $\theta =
\pi$. This can be circumvented by directly considering the time
dependence of $\rho$ at these points and will not affect the
following.  

When deriving the equations of motion for the new variables from
(\ref{eom_c1})-(\ref{eom_p}) one observes that only the difference
$q_2-q_1$ couples to the excitonic degrees of freedom. Therefore we
can introduce a dimensionless difference coordinate and the conjugate
momentum
\be
Q:= \sqrt{V}(q_2-q_1) \sep P:= {1\/2\sqrt{V}}(p_2-p_1)
\ee
and reduce the number of independent variables in this way by two.
The reduced equations of motion for the remaining five variables are obtained 
after the introduction of a dimensionless time 
\be\label{tau}
\tau = 2V\,t
\ee
and dimensionless parameters
\be
p:={\ga^2\/2V\,\om^2}\sep r:={\om\/2V}\sep g:={\Ga\/4V}
\ee
describing the strength of the electronic-vibronic coupling, the
frequency ratio of the two interacting subsystems and the strength of
the sink, respectively.
We find
\begin{eqnarray}
\label{EOM_R} {\dot R}&=& -g (\cos \theta + 1) R\\
\label{EOM_theta} {\dot \theta} &=& g \sin \theta + \sin \phi\\
\label{EOM_phi} {\dot \phi} &=& \cot \theta \cos \phi -  \sqrt{2p}\,r\,Q \\
\label{EOM_Q} {\dot Q} &=& P \\
\label{EOM_P} {\dot P} &=& -r^2\,Q-\sqrt{p/2}\,r\,R\,\cos \theta\,.
\end{eqnarray}
In these equations $({\dot{}})$ denotes ${d/d\tau}$. 

%%%%%%%%%%%%%%%%%%%%%%%%%%%%%%%%%%%%%%%%%%%%%%%%%%%%%%%%%%%%%%%%%%%%%%

\subsection{The DST Approximation}
One standard way to simplify the dynamics of excitonic-vibronic coupled
systems is to assume that the vibronic degrees of freedom
instantaneously adapt to the state of the excitonic subsystem and always
remain in the ground state prescribed by it. Applied to
(\ref{eom_c1})-(\ref{eom_p}), this assumption results in the DST
equations mentioned in the introduction. 

Within our effective dynamic model (\ref{EOM_R})-(\ref{EOM_P}) the DST
approximation can be justified assuming a separation of time scales. The
time scale for the (free) transfer of the excitation between the two
sites of the dimer has been normalized to 1 when the equations of motion
were written using the dimensionless time $\tau$ (\ref{tau}). Another
relevant characteristic time of our system is the period of the oscillator which
is $\sim 1/r$. Now we assume that the vibronic degrees of freedom are
much faster than the exciton, i.~e.\ $r\gg 1$. In this case the
oscillator coordinate completes many cycles during a time of the order
$1$ which is relevant for the slow excitonic subsystem.  Consequently,
dynamic self averaging over the fast oscillator coordinate occurs and
$Q$ in (\ref{EOM_phi}) can be replaced by its time average. For the
harmonic oscillator $(Q, P)$ this average is given for arbitrary
amplitude by the oscillator ground state which is determined by the
state of the excitonic subsystem and which represents at the same time
the only fixed point of the oscillator dynamics, i.~e.\ formally we can
introduce the DST approximation by requiring quasistationarity in the
vibronic variables ${\dot Q} = {\dot P} = 0$. We obtain for the mean 
value of $Q$
\be\label{QDST}
Q_{\mrm{DST}}=-{1\/r}\sqrt{p\/2}R\cos\theta\,.
\ee
and substitute it  into (\ref{EOM_phi}) which is replaced by
\be
\label{EOM_phi_DST} {\dot \phi} = \cot \theta \cos \phi - p\,R\, \cos
\theta\,.
\ee
Together with (\ref{EOM_R}) and (\ref{EOM_theta}) this equation
governs the DST dynamics of our model.

An analytical justification for the averaging procedures applied in
this and the following sections can be given using mathematical tools
that were developed in the theory of nonlinear differential equations
(see e.~g.\ \cite{Ver90}) and will not be discussed here.  Instead we
confirm the resulting equation (\ref{EOM_phi_DST}) by the numerical
simulations for $r\gg 1$ presented at the end of this paper.

The derivation of the DST approximation using fast oscillator dynamics
is questionable although it seems straightforward within the mixed
quantum-classical description to which this paper is confined.
However, the assumption $r\gg 1$ means that the mixed
quantum-classical description itself looses its justification and
should be replaced by a full quantum treatment.  More consistent ways to
address the validity of the DST limit are based on dissipation
due to a quantum heat bath \cite{VAG94} and therefore beyond the scope of
our model.

The DST equation is known to reproduce at least qualitatively some
remarkable features which the full dynamic system \cite{ES} displays
for an arbitrary value of $r$, e.~g.\ the bifurcation in the phase
space for $p = 1$ \cite{ELS85}, the resulting possibility of self
trapped solutions for coupling strengths above this value \cite{KC86}
and the possibility of dynamical chaos when an external perturbation
is applied \cite{EH91}. We can therefore consider the results obtained
in \cite{BES95} within the DST approximation for a dimer with sink as
a guiding line for the effects that can be expected in the present
more complete treatment.

\section{Quasistationary Decay Modes}
\label{3}
\subsection{Fixed Points for Quasistationary Total Occupation}
\label{3.1}
Beside the free exciton transfer time and the oscillator period a third
relevant time scale controls the decay of the excitation.  As we shall
see it is not necessarily given by the inverse sink rate $1/g$.  In the
present section we will assume that the exciton decay is the slowest
process in the system and discuss the resulting quasistationary
solutions of the equations of motion (\ref{EOM_R})-(\ref{EOM_P}). In the
special case of the DST limit these quasistationary solutions are most
transparent, while the opposite case of an oscillator which is even
slower than the excitation decay will be discussed in the next
subsection and results in different quasistationary solutions. 

The further analysis is motivated by the observation that a self
contained equation for the decay of the total occupation probability
$R(\tau)$ would be obtained if the function $\theta(R)$ was known.  This
is the case, in particular if the system remains in a fixed point of the
equations (\ref{EOM_theta})-(\ref{EOM_P}) while $R$ varies slowly enough
to be considered as an adiabatic parameter of these equations.  Then,
according to (\ref{EOM_R}), either the sink rate $g$ is very small or
the quasistationary state has a site occupation difference which is
strongly biased towards the site without sink $\cos\theta\sim -1$. 

In the following we discuss the fixed points of the equations
(\ref{EOM_theta})-(\ref{EOM_P}) with R treated as an adiabatic
parameter.  The equations (\ref{EOM_Q}) and (\ref{EOM_P}) yield for a
fixed point $P=0$ and $Q=Q_{\mrm{DST}}$, i.\ e.\ the location of the
fixed points is the same for the system with the full oscillator
dynamics included and for the DST equations. The equations
(\ref{EOM_theta}) and (\ref{EOM_phi}) allow for two different pairs of
fixed points on the Bloch sphere classified in what follows as
detrapped and self trapped states.  The stability exponents for
any of these points can be obtained from a linearization of the
equations of motion (\ref{EOM_theta})-(\ref{EOM_P}) around the fixed
point which yields the characteristic equation

\be
0=(\lambda^2+r^2)\([\lambda+\sin\phi\,\cot\theta][\lambda-g\,\cos\theta]+
{\cos^2\phi\/\sin^2\theta}\)-r^2p\,R\,\cos\phi\,\sin\theta\,.
\ee

\subsubsection*{(A) Detrapped states} 
For sufficiently small sink rate $g \le 1$ we obtain two
fixed points at
\be
\label{FPAPHI}
\sin \phi = - g, \hspace*{1cm} \cos \phi = \pm \sqrt{1 - g^2}
\ee
\be
\label{FPAT}
\cos \theta = 0\,.
\ee
Because of (\ref{FPAT})
the occupation probabilities for the two sites are the same and
we call the fixed points $A^\pm$ detrapped states. The point A$^+$ at $\cos\phi
> 0$ can be considered as a generalization of the bonding state in the
system without sink whereas the point A$^-$ at $\cos\phi < 0$ corresponds to 
the antibonding state. The stability exponents of these fixed points are
given by
\be
\label{FPAL}
\lambda^2 = -{r^2+\cos^2\phi\/2}\pm\sqrt{\({r^2-\cos^2\phi\/2}\)^2
+r^2\,p\,R\,\cos\phi}\,.
\ee
For A$^+$ the argument of the square root is always positive and the
pair of stability exponents corresponding to the negative sign in
(\ref{FPAL}) is purely imaginary. The other pair consists of two
imaginary or two real exponents with opposite signs such that the fixed
point A$^+$ is elliptic if $(pR)^2 + g^2 < 1$ and unstable hyperbolic
otherwise. 

For A$^-$ the stability is determined by the argument of the square root
in (\ref{FPAL}). If it is positive, the point is a stable elliptic center
and this is always the case in the adiabatic regime $r \to 0$, in the
opposite DST case $r \to\infty$ or for arbitrary parameters at large
times since then $R \to 0$.  Only for $p$ sufficiently large the
argument of the square root may temporarily be negative. The stability
exponents then acquire real parts with opposite signs and the point A$^-$
renders unstable hyperbolic. 

For the DST case $r \gg 1$ one pair of stability exponents which is
given by $\lambda = \pm ir$ corresponds to the fast oscillations
around the DST solution whereas the other pair of exponents 
\be\label{FPALdst}
\lambda^2 = \cos \phi (pR - \cos \phi)
\ee 
describes the stability of the DST solution itself.

We note, that the positions of the two fixed points $A^\pm$ do not depend on
$R$ and are thus constant in time. For the full system with $R$ time
dependent they represent therefore special time dependent states in
which the distribution of the excitation over the two sites is constant,
just the total occupation decreases exponentially at a rate $g$
\be
\label{sdm1}
{\dot R} = -g\; R\;.
\ee
When these fixed points are stable, a state prepared in their vicinity 
will remain there and decay at a mean rate $g$ with some oscillations
superimposed. 

\subsubsection*{(B) Self trapped states}
For a strong sink or strong nonlinearity $(pR)^2 + g^2 \ge
1$ there exist two other fixed points with biased site occupation
probabilities, i.\ e.\ self trapped states:
\be
\label{FPBT}
\sin \theta = {1 \over \sqrt{(pR)^2 + g^2}}, \hspace*{1cm} \cos
\theta = \pm \sqrt{1 - \sin^2 \theta}
\ee
\be
\label{FPBPHI}
\sin\phi = -g \sin\theta \hspace*{1cm} \cos\phi = pR \sin\theta
\ee
The existence of these two fixed points corresponds exactly to the range
of parameters for which the fixed point A$^+$ is hyperbolic and for
$(pR)^2 + g^2 = 1$ the points A$^+$ and $B^\pm$ merge into a single one. So we
have established a generalization of the pitchfork bifurcation of the
system without sink \cite{ES}. The difference is that the bifurcation
parameter no longer depends exclusively on the coupling strength $p$.
Rather it contains along with $p$ the strength of the sink $g$ and the
total occupation probability $R$ which is a function of time. If $g < 1$
the fixed points $B^\pm$ will disappear for large times when $R\to 0$ and we
can speak of a dynamic bifurcation. On the other hand, when $g > 1$ the
detrapped states A do not exist anymore and there is no bifurcation in
the course of time. 

Another crucial difference to the sink less dimer has to do with the
character of the points B. Without sink they are stable elliptic centers
and in their vicinity there exist solutions which remain self trapped
for all times. In the present case the stability exponents have to be
determined from the quartic equation
\be
\label{FPBL}
0=(r^2+\lambda^2)([\lambda-g\,\cos\theta]^2+p^2\,R^2)-(r\,p\,R\,\sin\theta)^2\,,
\ee
which does not allow for an easy solution. We shall see later
on, that the fixed points $B^\pm$ due to their time dependence do not  
represent quasistationary solutions unless the bifurcation parameter
$\sin^{-2}\theta=(pR)^2 + g^2$ is far enough above the bifurcation value 1.
Therefore we simplify (\ref{FPBL}) under the assumption $\sin^2\theta \ll
1$ and drop the second term. Then, there are two pairs of solutions for
the stability exponents. One of them is purely imaginary $\lambda = \pm
ir$ and the other one contains a real part as well
\be
\label{Lambda2}
\lambda = g \cos \theta \pm i\,pR\,.  
\ee
Again, for $r\to \infty$ the first pair describes the oscillations
around the DST solution and the other one the stability of the
DST solution which can alternatively be obtained from a
linearization of (\ref{EOM_theta}) and (\ref{EOM_phi_DST}) without further
approximations as
\be
\label{FPBLDST}
\lambda = g \cos \theta \pm i\,|\cos\theta|\,pR\,.  
\ee
Note that $\pm$ in (\ref{FPBT}) stands for the two
different fixed points, whereas the $\pm$ in (\ref{Lambda2}) and
(\ref{FPBLDST}) corresponds to two different stability exponents
$\lambda$ of the same fixed point.

The character of the fixed points $B^\pm$ is determined by the real part of
$\lambda$. For B$^+$ with an occupation bias towards the sink site
($\cos\theta > 0$) we have a repeller whereas the point B$^-$ with a low
occupation probability at the sink site ($\cos\theta < 0$) is a stable
attractor.

It is less clear than for the detrapped states that the fixed points B
which were obtained under the assumption of a constant total
occupation have some interpretation for the full system since their
location does depend on $R(\tau)$. However, we will show now that at
least B$^-$ can represent an attractor and a quasistationary solution
for the complete equations of motion (\ref{EOM_R})-(\ref{EOM_P}) when
we are sufficiently far away from its threshold of existence, i.\ e.\ 
when the bifurcation parameter is sufficiently large
\be
\label{STAB}
\sin^2 \theta = \frac{1}{g^2 + (pR)^2}\ll 1\,.
\ee
For this purpose we have to show that the change in the position of
B$^-$ is much slower than the relaxation towards this fixed point. 
The latter occurs on a time scale given by (\ref{Lambda2}) as $1/g$, while 
the oscillations around the fixed point have a period $2\pi/r$ and will
be averaged out provided this time is small enough. 

The velocity of the fixed point location can be estimated after
inserting (\ref{EOM_R}) into (\ref{FPBT}) and (\ref{FPBPHI}). We
obtain 
\begin{eqnarray}
|{\dot \theta}| &=& g (pR)^2 \frac{\sin^3\theta(1 +
\cos\theta)}{|\cos \theta|}\nonumber\\
&\sim& \frac{g(pR)^2}{2} \sin^5\theta \nonumber\\
&<& \frac{g}{2} \sin^3\theta 
%\ll |\Re\, \lambda|
\nonumber\\
|{\dot \phi}| &=& (gpR)^2 
{\sin^3\theta (1 + \cos\theta) \over \cos \phi}\nonumber\\
&\sim& \frac{g^2pR}{2} \sin^4\theta \nonumber\\
&<& \frac{g}{2} \sin^2\theta 
%\ll |\Re\, \lambda|
\nonumber
\end{eqnarray}
and conclude that the position of B$^-$ changes always slowly when
$\sin^2 \theta \ll 1$ and in particular the relaxation towards the
attractor is much faster. If moreover $g/2\,\sin^2 \theta \ll r$ the
oscillations around the fixed point average out in (\ref{EOM_R}) and we
can consider ($R(\tau)$, $\theta[R(\tau)]$, $\phi[R(\tau)]$) as an
attractor for the full system. This condition is automatically satisfied
when $r > 1$ (e.\ g.\ in the DST regime), but in the adiabatic regime $r
\ll 1$ the assumption of quasistationarity of B$^-$ may not be valid or
restricted to a short interval in time. In this case the following does
not apply and has to be replaced by the discussion in the next
subsection. 

Once the system is close to the attractor B$^-$ the excitation decay
is approximately governed by the equation
\be
\label{sdm2}
{\dot R} = -{g/2\/(pR)^2 + g^2} R\,.
\ee
If the sink rate is dominant $pR < g$ we obtain from (\ref{sdm2})
the interesting effect that the decay rate {\em decreases} for
increasing sink rate $g$. The reason is that the location of B$^-$ is
strongly shifted to the site without sink such that the probability to
find the exciton on the sink site becomes very small. This behavior
has previously been studied without coupling to vibrations and was
termed "fear of death" effect in some publications \cite{Bar91}. When
we have $pR/g \to 0$ which is always the case for $\tau \to \infty$
(\ref{sdm2}) implies an exponential decay of the exciton with the
rate ${1/2g}$. From (\ref{sdm1}) and (\ref{sdm2}) we conclude that the
fastest quasistationary decay of the excitation is realized for small
electronic-vibronic coupling and a sink rate $g=1$. 

Self trapping on the site with sink is not immediately destroyed when
the sink becomes effective. Though the state B$^+$ is a repeller, the
solution can oscillate with slowly increasing amplitude around this
fixed point provided that $g$ is not too large. In this case B$^+$ can
control the dynamics for some finite time which then leads to an
enhanced excitation decay.  We do not want to pursue this possibility
further, since we restrict our discussion to an excitation initially
localized outside the sink and in this case the point B$^+$ is
always irrelevant. 

To end this section we would like to recall some of the numerical
results obtained in \cite{BES95} for the DST dimer. There we found that
the system relaxes to a quasistationary self trapped state provided that
the nonlinearity was sufficiently large. Moreover, it was demonstrated
that the sink supported the tendency to self trapping and that the self
trapped state disappeared for large time when the sink was not too
strong. For strong sink we showed that the life time of the exciton was
the larger the stronger the sink was chosen. Now we are able to
interpret these qualitative observations as the dynamical manifestations
of the point B$^-$ and can give them a quantitative formulation using
the results of this subsection. 

\subsection{The Adiabatic Regime}
\label{3.2}
In the previous subsection we considered quasistationary dynamics of
the excitation under the assumption that the decay of the exciton
represents the slowest process in the system. This assumption breaks
down when the system is in the adiabatic regime $r \ll 1$.  Then
the decay is an essentially nonstationary process leading to the
disappearance of the exciton before the oscillator has gone through 
a large number of cycles.

Opposite to the derivation of the DST equation we can now assume that
the exciton completes many oscillations on the time scale of the
oscillator. Again we want to exploit this fact by averaging over the
variables of the fast subsystem and replacing them by their mean value, 
but in contrast to the harmonic oscillator of the DST case, the equations
of motion for the excitonic variables have two fixed points for a given $Q$
which are obtained by setting the 
l.~h.~s.\ of eqs.\ (\ref{EOM_phi}) and (\ref{EOM_theta}) to zero. This
results in
\be
\sin\phi= -g \sin\theta \sep \cos\phi=\sqrt{2p}\,r\,Q\,\tan\theta
\ee
and combining these two equations we find
\be
\label{FPAD}
g^2\,\sin^2\theta + 2p\,r^2\,Q^2\,\tan^2\theta=1\,.
\ee
The latter equation determines the location of the fixed points which depends
parametrically on $Q$.  The stability exponents can be
given in the form
\be\label{FPADL}
\lambda = g\cos\theta \pm i\sqrt{2p}\,r\,|Q/\cos\theta|\,.
\ee
% \be
% \cos\theta=\pm{1/\sqrt{2}g}\(\sqrt{(2pr^2Q^2+1-g^2)^2+8pg^2r^2Q^2}-
% (2pr^2Q^2+1-g^2)\)^{1/2}
% \ee
Eq.\ (\ref{FPAD}) is a biquadratic equation in $\cos\theta$, i.~e.\ the two
fixed points have opposite signs for $\cos\theta$. 
Due to (\ref{FPADL}) this means that one of them
is a stable attractor, the other one a repeller.
Instead of writing down the explicit solution of (\ref{FPAD}) 
which is quite a lengthy expression though easily found
we would like to mention two limiting cases.

First we note, that as $g\to 0$ the two fixed points approach the
well known lower and upper adiabatic states of the system without trap
(see e.~g.\ \cite{SE96}) which are given by
\be
\cos\theta=\pm{\sqrt{2p}\,r\,Q\/\sqrt{1 + 2p\,r^2\,Q^2}} \sep
\cos\phi=\pm 1\,.
\ee
The second  important limiting case corresponds to a strong localization of
the exciton on one of the two dimer sites ($\cos\theta\to\pm 1$). Assuming
$g^2 + 2p\,r^2\,Q^2 \gg 1$ the solution approaches
\be\label{FPADsl}
\sin\theta = {1\/\sqrt{g^2+2p\,r^2\,Q^2}}\,.
\ee
In particular, for a very large sink rate $g$ the dependence on $Q$
disappears and we have 
\be\label{FPADst}
\sin\theta={1\/g}\ll 1 \sep \sin\phi=-1\,.
\ee
In order to be able to replace the time average for the excitonic
variables by the discussed fixed points we have to assume that the time scale
given by (\ref{FPADL}) for the relaxation is sufficiently short, or
that the oscillations around the fixed point do average out. Due to the
nonlinear equations for the excitonic variables and in contrast to the
derivation of the DST equation we have in this latter case to assume
that the amplitude of the oscillations is small, i.~e.\ the system
has to be prepared close to one of the adiabatic states. If so, even 
the adiabatic state with $\cos\theta < 0$, which is in fact a repeller, can be
considered quasistationary for some limited time. 

A self contained equation for the decay of the total excitation
probability $R(\tau)$ can be derived under the assumption that the
exciton is located close to the attractive quasistationary state.  Then
$\cos\theta$ may be replaced by the value prescribed by the oscillator
coordinate $Q$ according to (\ref{FPAD}) and is constant provided the 
oscillator dynamics is sufficiently slow to be completely disregarded during
the life time of the exciton or if the position of the fixed point does
according to (\ref{FPADst}) not depend on time due to a strong trap.
Either case leads to an exponential decay of the excitation.
 
%%%%%%%%%%%%%%%%%%%%%%%%%%%%%%%%%%%%%%%%%%%%%%%%%%%%%%%%%%%%%%%%%%%%%%%
\section{The Time Evolution of the System}
\label{4}
\subsection{Parameter Regions and Initial Conditions} 
In the previous section we have discussed possible quasistationary
states of our system without indicating, if and when these states will
be reached by a particular solution of the equations of motion.  Now we
turn to the investigation of the time evolution of particular solutions
starting from specified initial conditions. 

As mentioned in the introduction we will always consider an exciton which is 
created at the site without trap, i.~e.\ we set initially
\be
c_1(0) = 1 \sep c_2(0) = 0\,.
\ee
The initial conditions for the vibronic degrees of freedom which we have
chosen are meant to take into account different physical possibilities
to prepare the excitation and to provide enough variety to estimate the
degree to which the excitonic variables depend on the details of the
oscillator initial state. The resulting solutions will be referred to in
the following way:
\begin{enumerate}

\item {\em Bare exciton}:
The first initial state we consider corresponds
to a creation of the exciton
on the first molecule by a very short light pulse.
During the optical excitation from the ground state there is not enough time
for the local vibrations to accommodate to the creation of the exciton.
This means we assume the vibrations initially in their ground
state {\em without} exciton 
$q_{i}(0) = 0$, $p_{i}(0)=0$.
The total energy for this initial condition is 0.

\item {\em Polaron}:
The second possibility is to assume a slow excitation such that
initially the vibrational degrees of freedom are already relaxed to
their new ground state with exciton. This is the initial condition which
would be implied by the DST approximation: 
$q_1(0)=-\ga/\om^2$, $q_2(0)=0$, $p_{i}(0)=0$.  
%$Q(0)=Q_{\mrm DST}=\sqrt{p/2}/r$, $P(0)=0$ 
The total energy of the polaron is $-{\ga^2/2\om^2}$, i.~e.
lower than the energy of the bare exciton. 

\item Polaron with additional vibrational energy [{\em Polaron (--)} and
  {\em Po\-la\-ron (+)}]:
  The different initial energies make a direct comparison between the
  bare exciton and the polaron difficult.  Therefore we have taken
  into account a third possibility for the initial condition which is
  not directly related to a particular way of preparing the exciton.
  Again we chose the configuration coordinate of the vibrations in the
  minimum of the potential after the exciton has been
  created. We supply, however, an initial momentum such that the total
  energy is 0 as for the bare exciton case. For this momentum we have
  two different possible directions i.~e.\ the polaron ($\pm$) is
  specified by 
  $q_{1}(0) = -\ga/\om^2$, $q_2 = 0$, $p_1(0) = \pm\ga/\om$, $p_2(0) = 0$.

\end{enumerate}
Before we explore the different types of solutions using the results
of section \ref{3} as well as numerical simulations we would like to 
briefly mention some limiting cases for
which analytical solutions to our problem can be given.  For a
sinkless DST dimer ($g=0$) exact solutions in terms of elliptic
functions were found in \cite{KC86}. Starting from a state completely
localized on one of the dimer sites they turned out to be self trapped
for $p > 2$. Recently, for certain parameter regions and $g=0$ exact
solutions with the oscillator dynamics explicitly taken into account
in a mixed quantum-classical description were also obtained
\cite{KK94a}.

For the linear dimer ($p=0$) with a sink the equations of
motion can be integrated and yield for the total occupation
probability
\be
\label{ldim}
R(\tau) = \left(
1 - G^2 + \frac{G^2 - G}{2}e^{+\sqrt{g^2 - 1}\tau}
        + \frac{G^2 + G}{2}e^{-\sqrt{g^2 - 1}\tau}
\right)e^{-g\tau}
\ee
with 
$G \,:=\,\frac{g}{\sqrt{g^2 - 1}}\,.$
For the full set of equations of motion (\ref{EOM_R})-(\ref{EOM_P}) we
have analytical solutions for exceptional cases only such as the
quasistationary states $A^\pm$ of section \ref{3.1}. Even for the DST approximation
with sink no exact solutions are known in general.

We have performed a numerical integration of the coupled system of
equations (\ref{eom_c1})-(\ref{eom_p}) for the different described initial
conditions. We display results for the total occupation probability
$R(\tau)$ and the relative site occupation difference expressed by
$\cos\theta(\tau)$ 
for various values of the oscillator frequency ranging from
the high frequency (DST) limit in figs.~\ref{0.1_1_10}-\ref{3_3_10}
to the deeply adiabatic region in  fig.~\ref{3_3_0.001}.
We restrict the sink rate and the vibrational coupling to three 
representative cases:
\begin{itemize}
\item weak sink $g=0.1$ / weak coupling $p=1$:
fig.~\ref{0.1_1_10}
\item weak sink $g=0.1$ / strong coupling $p=3$:
figs.~\ref{0.1_3_10}, \ref{0.1_3_1}, \ref{0.1_3_0.1} and \ref{3_3_0.001} 
\item strong sink $g=3$ / strong coupling $p=3$:
figs.~\ref{3_3_10} and \ref{3_3_1} 
\end{itemize}
For each set of parameters the results for the different initial
conditions will be displayed in the same graph. They can be
distinguished by the different line shapes annotated e.~g.\ in
\rf{0.1_1_10}(a).

%%%%%%%%%%%%%%%%%%%%%%%%%%%%%%%%%%%%%%%%%%%%%%%%%%%%%%%%%%%%%%%%%%%%%%

\subsection{Time Evolution in the DST Approximation}
Using the results of section \ref{3.1} we can obtain a quite
satisfactory description of the time evolution in DST approximation
that agrees with our numerical findings reported in \cite{BES95}.  We
have to distinguish three different cases with respect to the
parameters $p$ and $g$:

\subsubsection*{(I) Nearly Linear Regime $g^2 + p^2 < 1$}
In this case throughout the whole time evolution the only fixed points
present are the stable elliptic centers A from section \ref{3.1}.  The
relative site occupation difference $\cos\theta$ will therefore
oscillate with a mean value $\cos\theta = 0$.
The decay of the total occupation probability is then approximately given by
(\ref{sdm1}):
\be
\label{PI}
R(\tau) \sim e^{-g\tau}\,,
\ee
and is therefore very much like in the case of the linear dimer (\ref{ldim}). 

An illustration for the described behaviour is provided by
\rf{0.1_1_10}, which is with $p=1$ and $g=0.1$ at the fringe of
region (I).  Since the transition between the parameter regions
is smooth we find in \rf{0.1_1_10}(a) a straight line indicating
an exponential decay with some oscillations superimposed.
The mean decay rate obtained from the figure is in good correspondence to
(\ref{PI}) very close to $0.1$ and the period of the oscillations
in \rf{0.1_1_10}(b) is very close to $2\pi$. This is the value for the
free transfer of the excitation and corresponds to the asymptotic value for the
stability exponent of the points $A\pm$. However, the solution is actually
not in the vicinity of one of these points. Rather it oscillates with a large amplitude
and can therefore not be expected to be correctly described by a linearization
around a fixed point. For instance, the time dependence of the stability exponent 
(\ref{FPALdst}) is not reflected in the solution.

\subsubsection*{(II) Weak Sink $g < 1$ and Strong Coupling $g^2 + p^2 > 1$} 
In this case there exists the attractive fixed point B$^-$ from section
3.1 when the system starts its evolution at $R(\tau=0) = 1$.  As a
numerical example consider the DST curve of \rf{0.1_3_10} which is the 
thick gray line.  The system approaches the attractor after a time $\tau
\sim 1/g$ and then decays on it according to
(\ref{sdm2}). This is in general a nonexponential decay which is very
much different from (\ref{PI}).  If we assume strong nonlinearity $p \gg
1$ (\ref{sdm2}) can be approximated and results in
\be
\label{PII}
R(\tau) = \sqrt{1 - {g\over p^2}\tau}\;.
\ee
There will be oscillations around this mean behavior with an amplitude
decreasing as the attractor is approached. The frequency $\Omega$ of these
oscillations is given by the imaginary part of the stability exponent in
(\ref{FPBL}) and decreases approximately as $\Omega \sim pR$ for strong
nonlinearity. Indeed, the oscillations around the mean in \rf{0.1_3_10}(b)
have a period which can be seen to increase starting from $T\sim 2.4$ which
is close to the value $2.2$ obtained from the imaginary part of (\ref{FPBL}).
Then the  oscillations die out at $\tau\sim 30$ thus confirming the attractive
character of the fixed point B$^-$.

When the total occupation has decreased such that $g^2 + (pR)^2 \sim
1$, the attractor B$^-$ does not exist anymore and the system will
start oscillating with equal mean site occupation probabilities
around one of the fixed points $A^\pm$ as in (I).  The
time $\tau_0$ for the crossover from the algebraic decay (\ref{PII})
to an exponential behaviour with decay rate $g$ is approximately given by
\be\label{tau0II}
\tau_0 = \frac{p^2 - 1}{g}\,.
\ee
Of course the crossover does not occur instantaneously, rather there
exists a time interval close to the $\tau_0$ where we do not have a
good description for the system. We recall that the assumption of a
quasistationary decay on the point B$^-$ required the system to be
sufficiently far away from the dynamical bifurcation (\ref{STAB}).
This means that the crossover will actually start a little earlier
than predicted by (\ref{tau0II}) as can be seen in \rf{0.1_3_10} where
the crossover according to (\ref{tau0II}) should be at $\tau=80$ but
occurs actually at $\tau \sim 70$. It means also that the dynamics for
$g^2 + p^2 \sim 1$ will be hardly different from the case (I), where
an example for this transition behavior was already described with
\rf{0.1_1_10}.

\subsubsection*{(III) Strong Sink $g > 1$}
Here the attractor B$^-$ does exist throughout the evolution of the
system. If the nonlinearity is very large there might be initially a
nonexponential behavior as in (II), but this will
turn into an exponential decay as soon as the total occupation has
decreased sufficiently for $g \gg pR$. Then one has from (\ref{sdm2})
\be
\label{PIII}
R(\tau) \sim e^{-(g - \sqrt{g^2 - 1})\tau}\,.
\ee
Under the assumption $p\gg g$ which is, however, not satisfied in the
numerical example \rf{3_3_10}, the approximate crossover time is
obtained from (\ref{PII})as
\be
\tau_0 = \frac{p^2 - g}{g}\,.
\ee
The DST solution represented by the thick gray line in \rf{3_3_10}
relaxes after a very short time to the attractor $B^-$ whose initial
and final position is marked by dotted horizontal lines. Due to the 
larger $g$ compared to \rf{0.1_3_10} the initial oscillations around B$^-$ 
can hardly be observed. The crossover to constant relative site occupation 
probabilities and exponential decay occurs at $\tau\sim 20$.

We note that in all the three cases we discussed the limiting behavior
for large time $\tau$ and small total occupation $R$ agrees with that
of the linear dimer as given by (\ref{ldim}), i.~e.\ it is
exponential.  This is natural, since the second term in the DST
equation (\ref{EOM_phi_DST}) which due to the excitonic-vibronic
coupling is then negligible, but it is in contrast to the full system
of coupled excitonic-vibronic equations (\ref{EOM_R})-(\ref{EOM_P})
from which it is obvious that the oscillator keeps influencing the
dynamics of the exciton for all times. Just the feedback from the
exciton disappears as it decays.

\subsection{Deviations from the DST approximation at large but finite
oscillator frequency}
The three different scenarios for the time evolution in DST
approximation which were described in the previous section remain valid for the
full system with the oscillator frequency not too low, since
the fixed points of \ref{3.1} still represent quasistationary decay modes.
Using the numerical results displayed in \rf{0.1_1_10}--\ref{3_3_1} we
will demonstrate this, discussing at the same time deviations from the
DST solutions.

The degree of deviations from the DST solution will depend on the 
value of the parameter $r$ and on the initial conditions for the
oscillator. In particular for intermediate oscillator frequency it can
be expected that the DST approximation describes the actual solution
the better, the closer to it the initial condition for the oscillator
is chosen. Indeed, the polaron (0) (dashed thick gray line), which is
prepared in a DST state, cannot be distinguished at all from the DST
curve in the plots for high oscillator frequency $r=10$
(\rf{0.1_1_10}--\ref{3_3_10}) and follows it very closely in the plots
for the intermediate frequency $r=1$ (\rf{3_3_1}). For
small nonlinearity $p=1$ (\rf{0.1_1_10}) the same holds true for the
other three solutions which are prepared with higher energy than the
DST solution and in this case the initial conditions have no crucial
influence on the dynamics down to the intermediate oscillator
frequency $r=1$ (not displayed).

A systematic, though small deviation from the DST solution can be
observed for stronger nonlinearity in the figs.~\ref{0.1_3_10},
\ref{3_3_10} and \ref{3_3_1}. Here, a self trapped state is approached
by the polaron($\pm$) and the bare exciton solution as described for
the DST case, but beside the familiar slowly decaying oscillations we
find also oscillations of higher frequency which do not disappear
completely.  This behavior was to be expected from the stability
analysis of the fixed point B$^-$ for the complete system, where we
found from (\ref{FPBL}) beside the stability exponents of the DST
solution a pair $\lambda=\pm ir$ describing fast oscillations.
However, the location of the fixed point and the center of the
oscillations in figs.~\ref{0.1_3_10}, \ref{3_3_10} and \ref{3_3_1} are
not exactly the same. The full dynamic model tends to oscillate around
a state which is even more localized than predicted and consequently
it decays slightly slower.  Moreover, the period of the fast
oscillations is of the order of $2\pi/r$ but does not quite agree with
this value and is actually close to half of it.

The reason for these deviations can be seen in the fact that a
linearization of the flow around the fixed point is justified for
small amplitudes only - a condition not matched by the vibronic
variables, since the oscillator was prepared in a state of high
energy.  Moreover, after $\tau \sim 20$ when the exciton has almost
ceased to exist a description using fixed points of the coupled
excitonic-vibronic system -- though it still provides a fairly good picture
-- seems to be counter intuitive since we have seen that the
interaction in this case is one way only: the freely moving oscillator
represents an external perturbation to the excitonic subsystem. We
will come back to this point when we discuss the adiabatic case in the
next section.  For the time being it is sufficient to note that the
deviations shrink for growing $r$ as it can be seen by comparing
figs.~\ref{3_3_10} and \ref{3_3_1} and have completely disappeared on
the scale of the plots for $r=100$ (not displayed).

Unlike in the figures discussed so far, a qualitative difference in
the behavior of the three solutions prepared with total energy 0 is
observed for $p=3$ and $g=0.1$ when they are compared to the low
energy DST and polaron (0) solutions.  In \rf{0.1_3_1}(b) beside the
latter two the bare exciton is shown for which the oscillations around
the initially existing self trapped state $B^-$ are so large that this state is
hardly recognized at all. It disappears at $\tau \sim 10$ and thus
much earlier than for the DST and polaron (0) case. The other two
polaron solutions which are not displayed resemble the bare
exciton.

This behavior can be understood from the fact that the system with
$g=0.1$ is very similar to the sink less case which has been shown to
be strongly chaotic for $r=1$ and $p$ above the bifurcation value 1
\cite{ES}. Due to the chaos, the system explores the energetically
accessible phase space very fast and this is reflected in the strong and
irregular oscillations of the relative site occupation leading to a
rapid exciton decay.

\subsection{Time Evolution in the Adiabatic Case}
The conclusion of section \ref{3.2} was that there is an
exponential decay once the system is close to the attractive adiabatic
state.  However, there are two important limitations to this
conclusion.

First, the initial state of the exciton has to be close to one of the
adiabatic states. We consider the initially completely localized state
of the exciton $\cos\theta=-1$, i.~e.\ the initial oscillator coordinate
should correspond to a strongly localized adiabatic state. According to (\ref{FPADst})
this is the case if $g^2 + 2p\,r^2\,Q(0)^2 \gg 1$.  For the bare
exciton $Q(0)=0$ this is the case for large sink rate $g$ and then the
total occupation will decay close to the linear dimer (\ref{ldim})
independent on the actual strength $p$ of the coupling. For the
polaron we have 
\be
g^2 + 2p\,r^2\,Q(0)^2 = g^2 + p^2 \gg 1
\ee
and the exciton can be close to an adiabatic state even for small sink
rate provided the coupling to the oscillator is strong enough. The 
total probability decays in this case as
\be\label{ADdec}
R(\tau) \sim e^{-{g/2\/g^2+p^2}\tau}
\ee
which is initially very close to what is predicted from the
quasistationary decay mode B$^-$ (\ref{sdm2}). The largest possible decay rate
is according to (\ref{ADdec}) $1/4p$ and it is realized for $p=g$.

The second condition for an exponential decay of the occupation
probability is that the oscillator dynamics is actually sufficiently slow to
be completely disregarded during the life time of the exciton.
According to (\ref{ADdec}) this means 
\be\label{ADexp}
r \ll {g/2\/g^2+p^2}\,,
\ee
but the restriction of the exciton to one of the adiabatic states
prescribed by the oscillator is justified whenever $r \ll 1$, and this can be a
much weaker condition. So if in the adiabatic regime the condition
(\ref{ADexp}) is not satisfied, no self contained equation for the
decay of the exciton is available. 

This is the situation in \rf{0.1_3_0.1}. The parameter $r=0.1$ is
sufficiently small for the application of the adiabatic approximation.
Consequently, in part (b) of the figure the polaron (--) solution can
be seen to follow the evolution of one of the adiabatic states, namely
the energetically lower state obtained from
the solution of (\ref{FPAD}).  Initially, some decaying oscillations
around the adiabatic state can be observed which are in good agreement
with the stability exponents (\ref{FPADL}). When the adiabatic state
enters the region $\cos\theta>0$ it becomes a repeller and one
observes increasing oscillations around it until the variable $Q$ has
completed one full period at $\tau\sim 80$ and the relaxation to the
attractor starts again. 

The same behavior can be observed for the other two polaronic
solutions. The excitation decays rapidly as soon as the exciton is
driven by the oscillator to the sink site.  Therefore in this case the
life time of the excitation is basically determined by the frequency of the oscillator
and its initial conditions. Since the polaron (--) has an
initial momentum which is directed towards increasing polarization,
the exciton remains for a long initial period localized on the site
without sink. This period is shorter for the polaron (+) which 
has a momentum towards decreasing polarization and  consequently the
polaron (--) has a longer life time. The polaron (0) has no  
momentum at $\tau=0$ and decays initially at a rate in between the
other two polarons. Due to the lacking vibrational energy 
the oscillator coordinate for the polaron (0) changes its position only
very slowly such that the polaron (0) is the longest living solution.

Comparing in \rf{0.1_3_0.1}(a) the polaron (0) to the exciton we find
a good agreement up to the crossover time for the DST solution. Then
the DST exciton can be seen to decay faster than the polaron (0). The
reason is that the initially localized state of the exciton has for
the two solutions different sources. For the DST solution it results
from self trapping on the attractive fixed point B$^-$ which keeps the
exciton localized as long as it is far from its threshold of
existence. Once this threshold is reached, the exciton becomes
delocalized for good.

In contrast, for the polaron solution the oscillator is too inert to
change its position and thus keeps the exciton in a fixed adiabatic
state. While $B^-$ represents a unique point on the Bloch sphere for
given parameters and total occupation, there is nothing special about
the adiabatic self trapped state. In fact the exciton could for any
set of parameters be fixed anywhere if the initial condition for the
oscillator was chosen appropriately.  Moreover, as the oscillator
changes slowly its position, there will always be intervals when the
exciton is localized on the sink site, i.~e.\ the initial localization
of the exciton is {\em not} due to self trapping and the initial
agreement of the DST solution and the polaron (0) is simply due to the
fact that the system was prepared in a DST state.

This interpretation is further confirmed by the observation that the initial
agreement between the DST and the polaron (0) solution ceases to exist
as soon as the parameters do not support a self trapped state for the
DST case. The polaronic solutions are unaffected by this and do still
display an initial tendency towards localization on the sink less site
(not displayed).  Among them the polaron (+) again decays fastest while 
the polaron (--) is the longest living solution. 

The situation is similar in \rf{3_3_0.1}. The DST exciton relaxes due
to the strong sink quite fast to the final location of the fixed point
B$^-$ and then decays without further oscillations while the polaron
(0) due to its inertness remains for a longer time close to its
initial position and decays consequently slower than the DST solution.
All the polarons as well as the bare exciton keep oscillating around a
mean value which is slightly below the location of the fixed point
B$^-$. In fact \rf{3_3_0.1}(b) looks very much like the plots for
$g=3$ and $p=3$ at high and intermediate frequency \rf{3_3_10}(b) and
\rf{3_3_1}(b), just the deviation from the location of the point B$^-$
is larger and the oscillations are slower. But now we can provide a
more satisfactory explanation for this behavior using the adiabatic
states.  The polarons as well as the bare exciton follow after a very
short relaxation the adiabatic state at $\cos\theta<0$. As an example
for this behavior the adiabatic state for the polaron (--) is
displayed in \rf{3_3_1}(b) with sparse fat dots. The location of the
adiabatic state can be seen from (\ref{FPADsl}) to depend on the
squared amplitude of the oscillator coordinate. The localization is
weakest for $Q=0$, when the adiabatic state coincides with the point
B$^-$. The quadratic rather than linear dependence on $Q$ is the
reason why the adiabatic oscillations in the relative site occupation
observed in \rf{3_3_1}(b) have a mean value below $B^-$ and a
frequency which is exactly half that of the oscillator.

In contrast to all the other solutions, the bare exciton remains
completely unaffected by the nonlinearity in the adiabatic case.
Here, the oscillator is prepared at $Q=0$ and there it stays during
the whole life time of the excitation provided the adiabatic parameter
$r$ is sufficiently small.  Consequently the vibronic coupling has no
effect on the exciton and it oscillates independent on $p$ around
$\cos\theta=0$ for $g<1$ (\rf{0.1_3_0.1}) or relaxes to
$\cos\theta=\sqrt{1-1/g^2}$ otherwise (\rf{3_3_0.001}).  When the
coupling parameter $p$ is small, the bare exciton is well approximated
by the DST solution.  In the adiabatic regime $r\ll 1$ the bare
exciton shows among the different considered solutions at least
initially the fastest decay.

Finally we would like to discuss an example in which the condition
(\ref{ADexp}) for an exponential decay of the polaron solutions is
satisfied (\rf{3_3_0.001}). In this case the polarons do not
differ very much from each other and clearly follow an exponential
law at a rate very close to that predicted by (\ref{ADdec}). Since we
chose $p=g$ for the figure, the life time of the polarons is exactly
twice that of the bare exciton. The little remaining difference
between the polaronic solutions reflects the residual change in the
oscillator position during the life time of the excitation which
enhances the localization of the exciton for the polaron (--) and
diminishes it for the polaron (+) while there is no such effect for
the polaron (0).

%%%%%%%%%%%%%%%%%%%%%%%%%%%%%%%%%%%%%%%%%%%%%%%%%%%%%%%%%%%%%%%%%%%%%%

\section{Conclusions}

We have studied the decay of an exciton coupled to polarization
vibrations on a dimer. Quasistationary decay modes were identified
which allow to explain the basic properties of the system. Using
numerical simulations the deviations from the predicted behavior were
investigated.

The model exhibits a rich variety of dynamical regimes depending on
the parameters and the initial conditions.  We found effects such as
the time dependent bifurcation and the associated crossover in the
decay regime which are genuinely due to the interplay between the sink
and the vibrational coupling and cannot be explained by considering
one of these mechanisms alone.

The tendency to form an initially localized exciton state on the site
without sink is enhanced by both, vibrational coupling and trapping
due to the sink. For high and intermediate oscillator frequency the
system changes its behavior profoundly when the threshold for an
initially self trapped state is reached, while there is no such effect
in the adiabatic regime.

The relation between the DST approximation and a mixed
quantum-classical description, taking the oscillator dynamics
explicitly into account, was clarified. For high oscillator frequency
the influence of the oscillator initial condition is weak and the two
models behave very much the same. In the adiabatic regime the bare
exciton is close to the DST solution provided that the coupling is
weak.

The strong dependence on the initial conditions in the adiabatic case
make a careful description of the exciton creation process
indispensable for a satisfactory description of the system. Other
interesting possibilities to extend the model are the inclusion of
dissipation and / or quantum fluctuations.

\section{Acknowledgements}
Support from the Deutsche Forschungsgemeinschaft (DFG) is gratefully
acknowledged.  One of us (I.B.) has also enjoyed support from the
project GAUK 105/95 and from the Deutscher Akademischer
Austauschdienst (DAAD). While preparing this work, the authors
experienced the kind hospitality of the Humboldt University Berlin and
the Charles University Prague during mutual visits.

%\bibliographystyle{unsrt}
%\bibliography{twolevel,exciton,chaos}

%%%%%%%%%%%%%%%%%%%%%%%%%%%%%%%%%%%%%%%%%%%%%%%%%%%%%%%%%%%%%%%%%%%%%%

\clearpage
\vspace*{4cm}\begin{figure}
\centerline{\psfig{figure=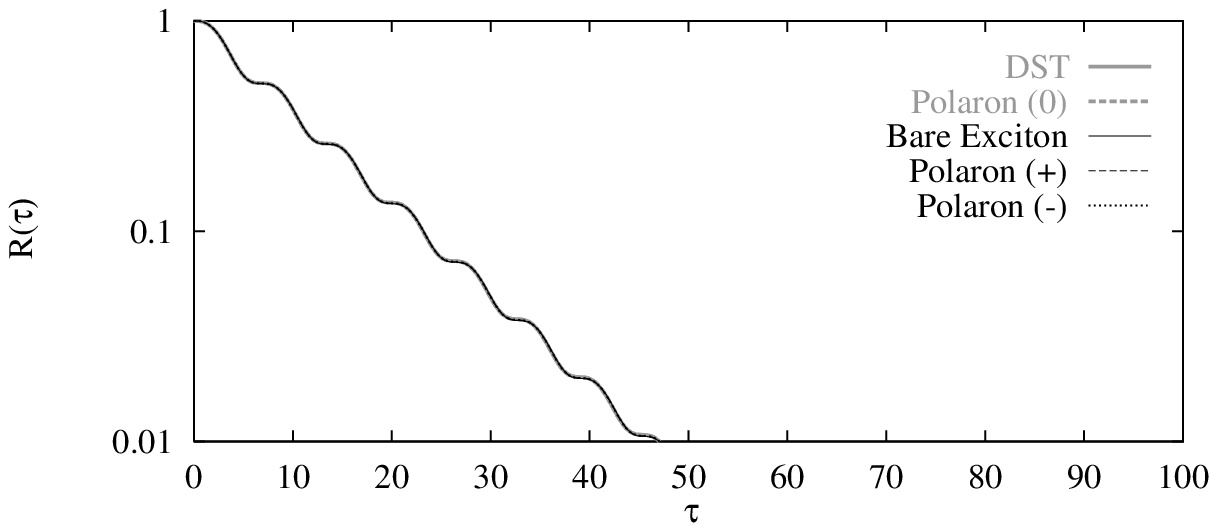,height=6cm,angle=0}}
\centerline{\psfig{figure=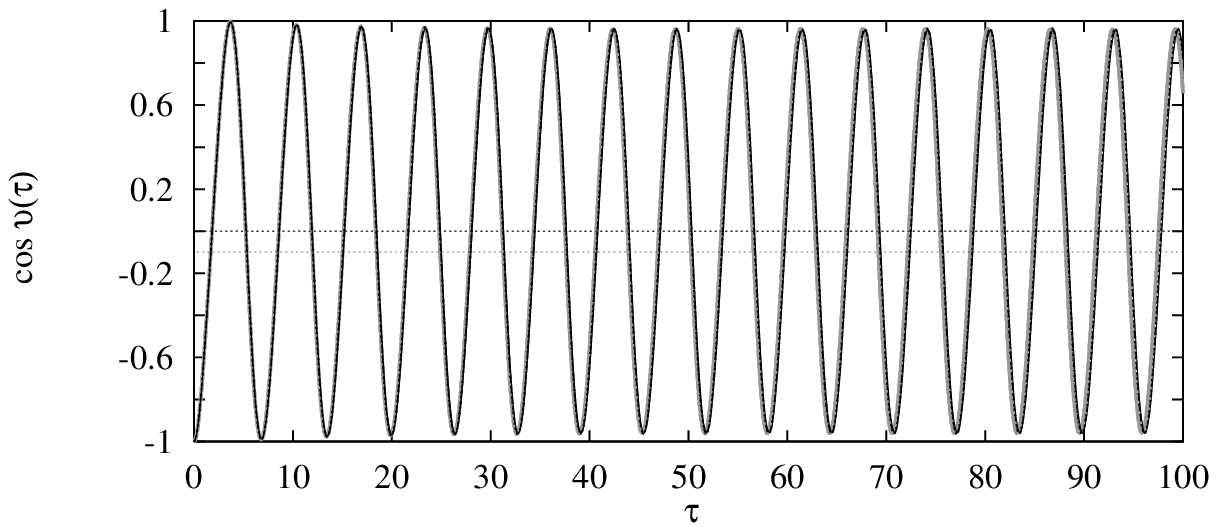,height=6cm,angle=0}}
\caption{
Time dependence of the total occupation (top) 
and of the relative site occupation difference (bottom)
for $g=0.1$, $p=1$ and $r=10$.
Different oscillator initial conditions for the full dynamic model 
and the DST dynamics can hardly be distinguished for this parameter set.
They are shown in this and all the following figures with the line types indicated
in the upper part. In the bottom plot, $\cos\theta=1$ corresponds to
the sink site and $\cos\theta=-1$ to the sinkless site where the exciton
is created. The lower/upper horizontal line 
shows the location of the fixed point B$^-$ at the time of the creation
of the exciton ($R=1$) and after its complete decay ($R=0$), respectively. 
}
\label{0.1_1_10}
\end{figure}

\clearpage\vspace*{4cm}\begin{figure}
\centerline{\psfig{figure=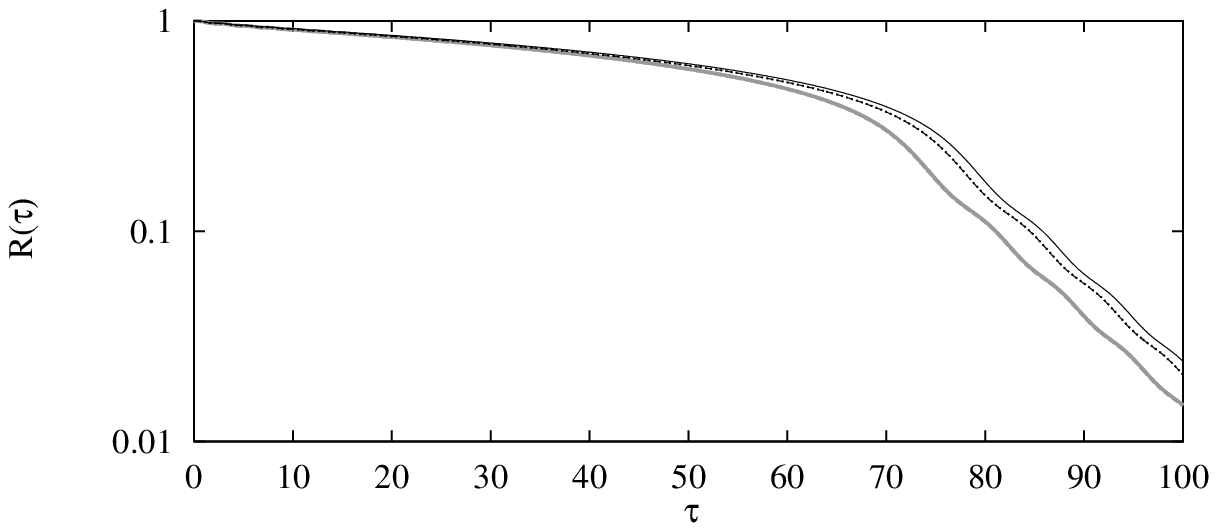,height=6cm,angle=0}}
\centerline{\psfig{figure=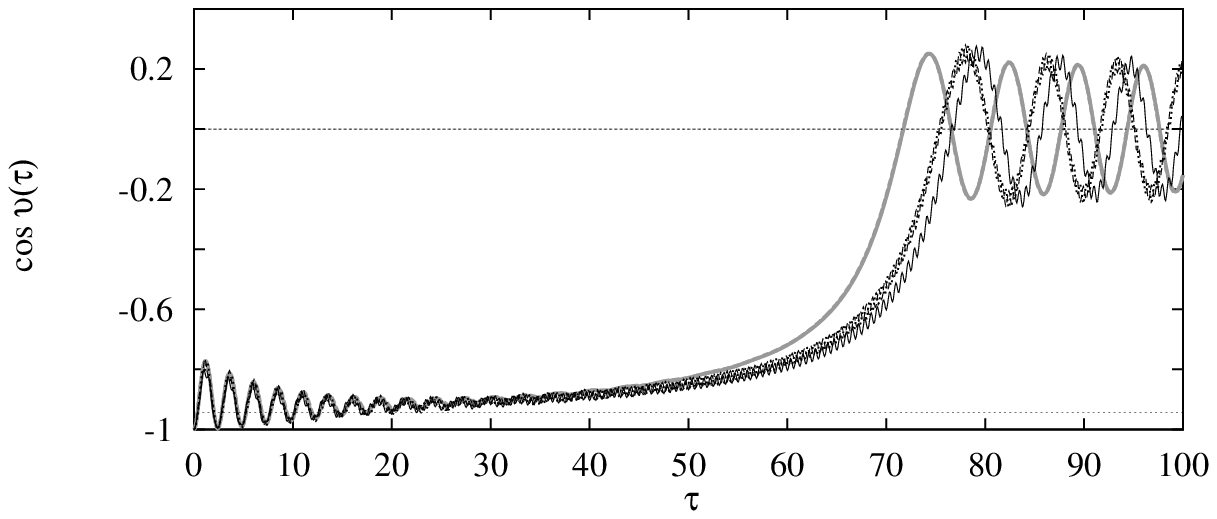,height=6cm,angle=0}}
\caption{
Time dependence of the total occupation (top) 
and of the relative site occupation difference (bottom)
for $g=0.1$, $p=3$ and $r=10$.
}
\label{0.1_3_10}
\end{figure}

\clearpage\vspace*{4cm}\begin{figure}
\centerline{\psfig{figure=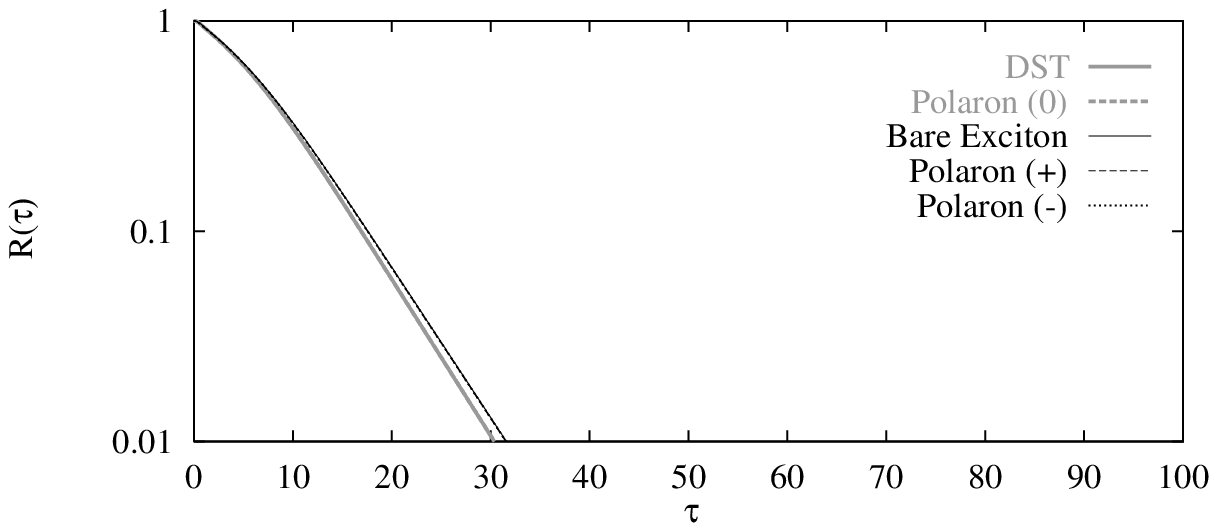,height=6cm,angle=0}}
\centerline{\psfig{figure=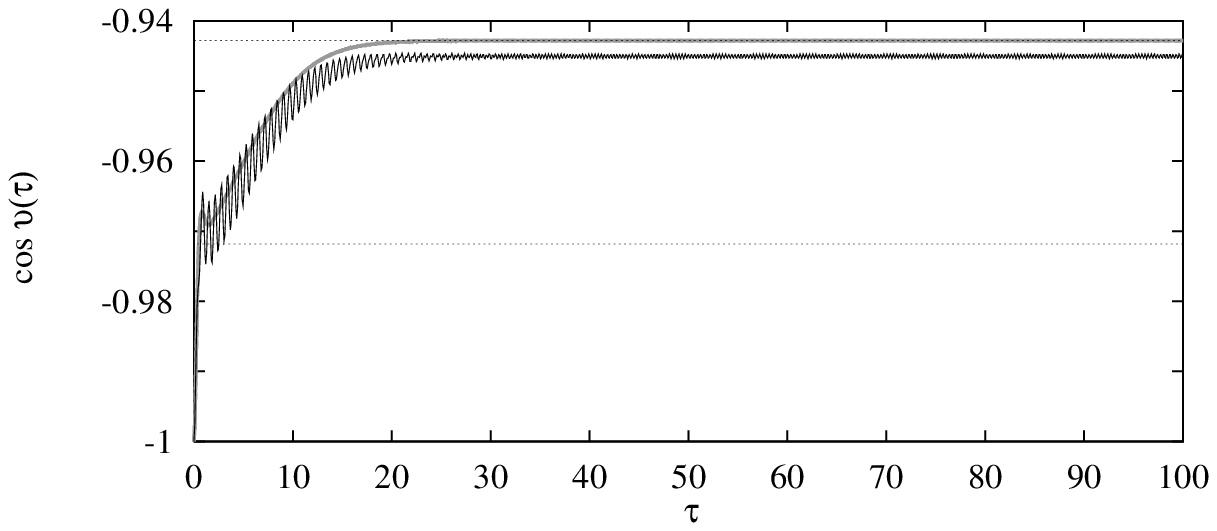,height=6cm,angle=0}}
\caption{
Time dependence of the total occupation (top) 
and of the relative site occupation difference (bottom - DST, 
bare exciton and polaron (0) only)
for $g=3$, $p=3$ and $r=10$.
}
\label{3_3_10}
\end{figure}

\clearpage\vspace*{4cm}\begin{figure}
\centerline{\psfig{figure=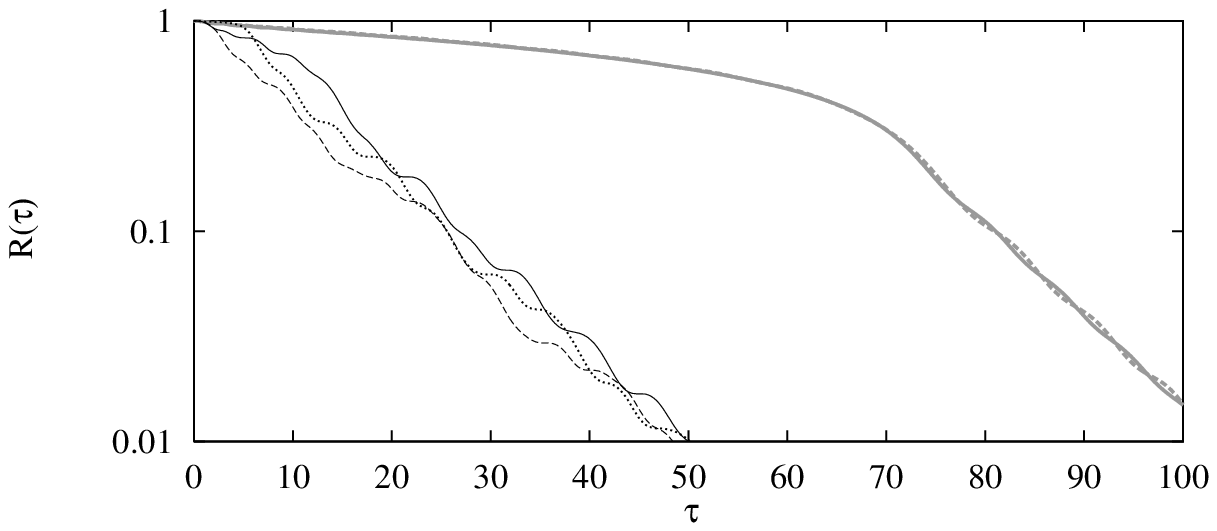,height=6cm,angle=0}}
\centerline{\psfig{figure=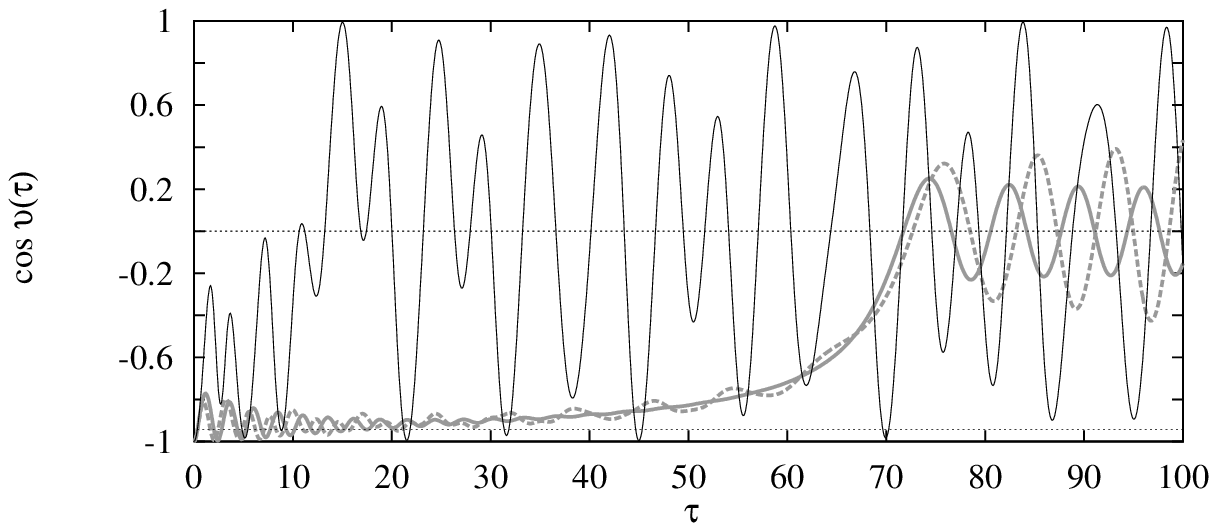,height=6cm,angle=0}}
\caption{
Time dependence of the total occupation (top) 
and of the relative site occupation difference (bottom - 
DST, bare exciton and polaron (0) only)
for $g=0.1$, $p=3$ and $r=1$.
}
\label{0.1_3_1}
\end{figure}

\clearpage\vspace*{4cm}\begin{figure}
\centerline{\psfig{figure=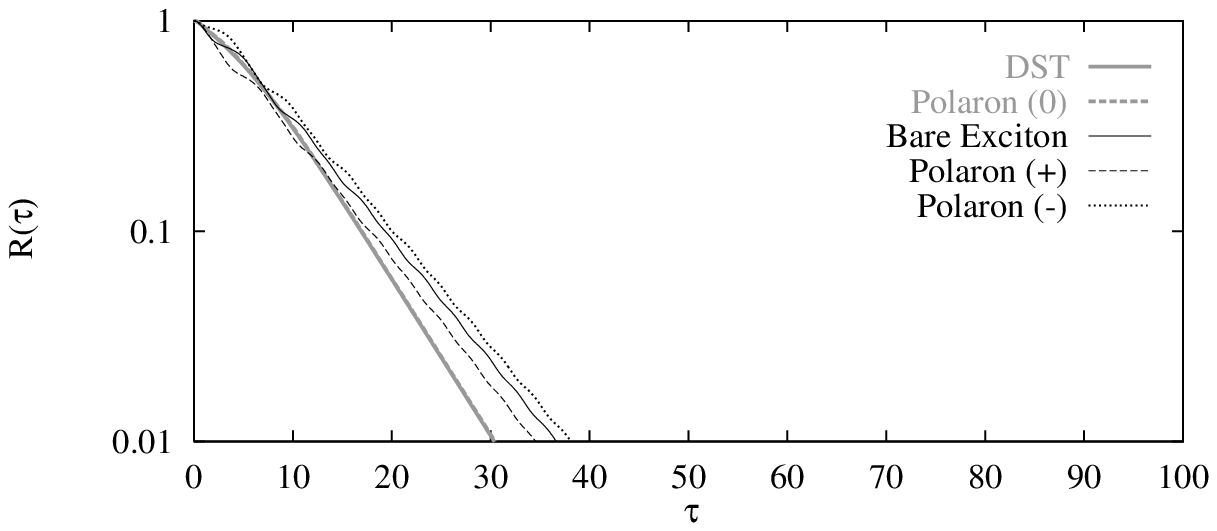,height=6cm,angle=0}}
\centerline{\psfig{figure=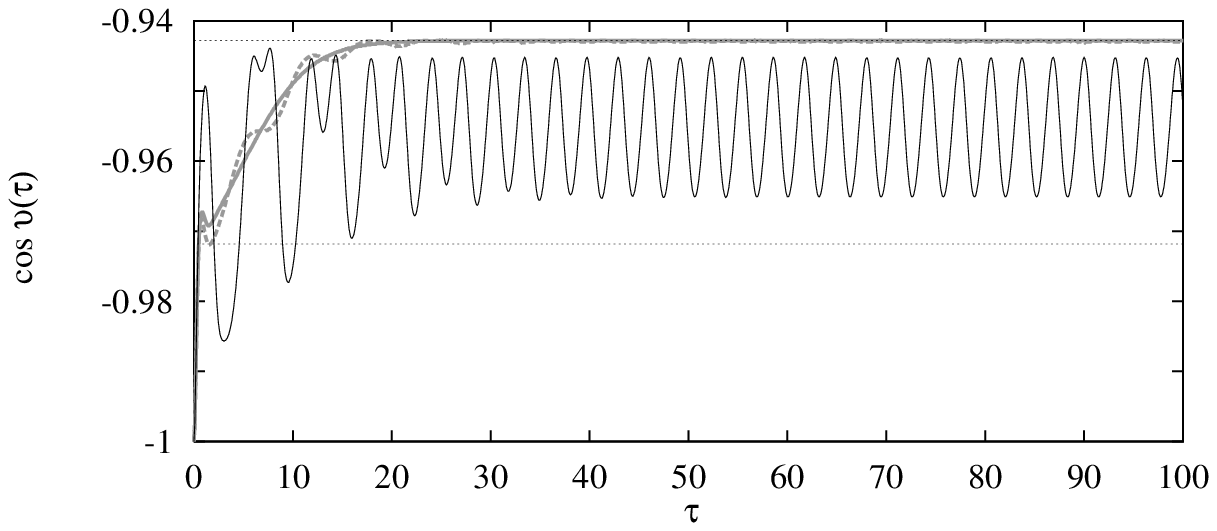,height=6cm,angle=0}}
\caption{
Time dependence of the total occupation (top) 
and of the relative site occupation difference (bottom -
DST, bare exciton and polaron (0) only)
for $g=3$, $p=3$ and $r=1$.
}
\label{3_3_1}
\end{figure}

\clearpage\vspace*{4cm}\begin{figure}
\centerline{\psfig{figure=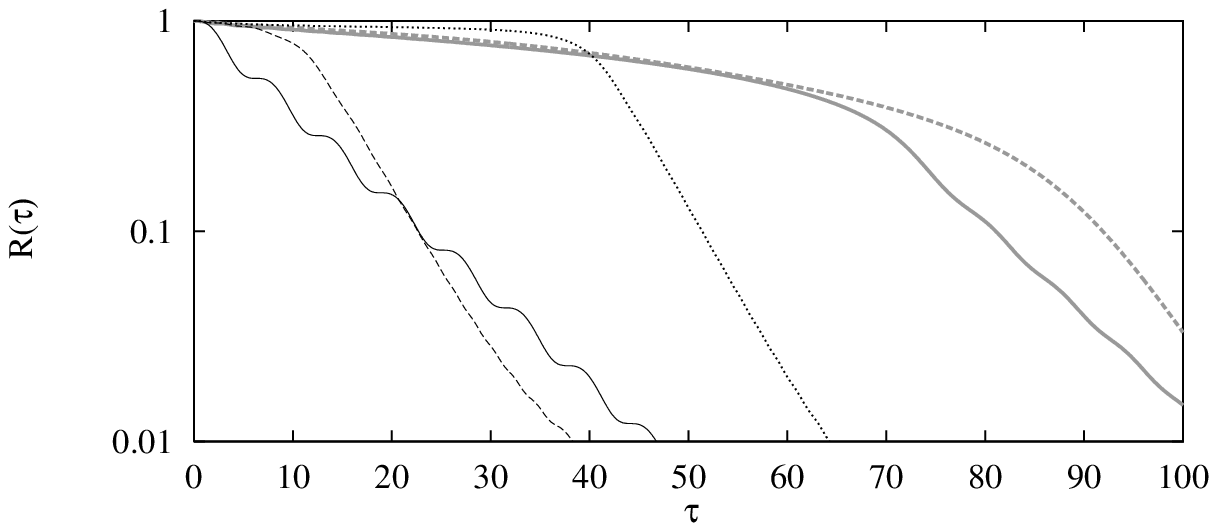,height=6cm,angle=0}}
\centerline{\psfig{figure=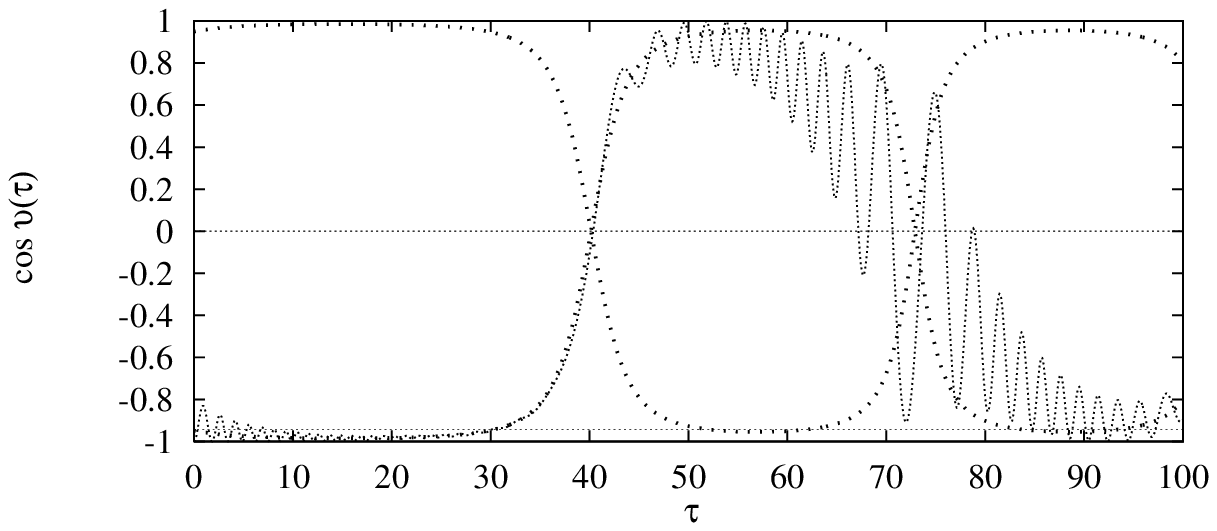,height=6cm,angle=0}}
\caption{
Time dependence of the total occupation for all initial conditions (top) 
and of the relative site occupation difference for the polaron (--) 
(bottom). With the sparse bold dots the time dependence
of the two adiabatic states is indicated.  
The parameters are $g=0.1$, $p=3$ and $r=0.1$.
}
\label{0.1_3_0.1}
\end{figure}

\clearpage\vspace*{4cm}\begin{figure}
\centerline{\psfig{figure=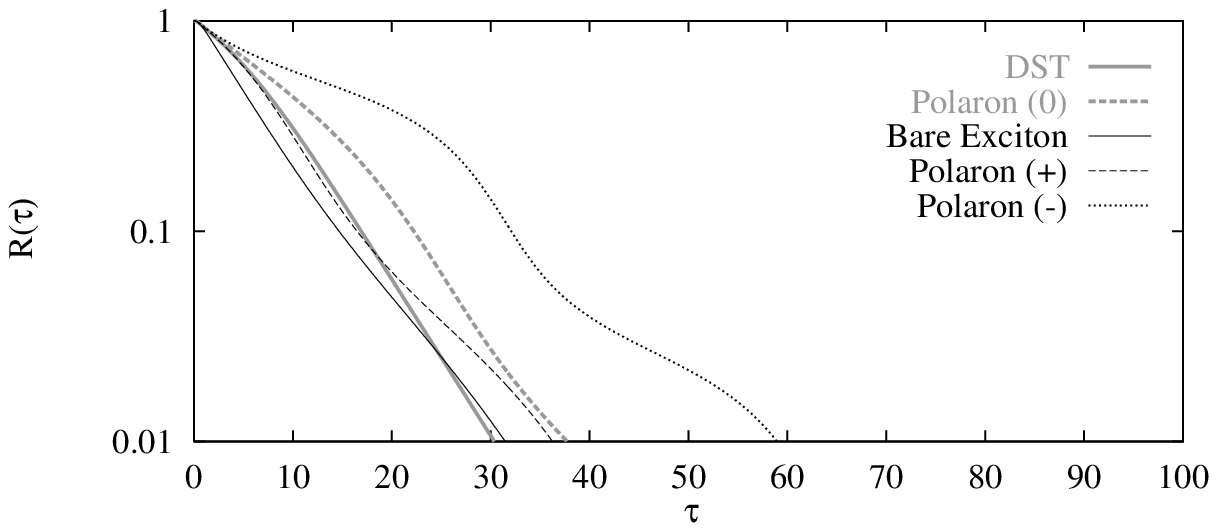,height=6cm,angle=0}}
\centerline{\psfig{figure=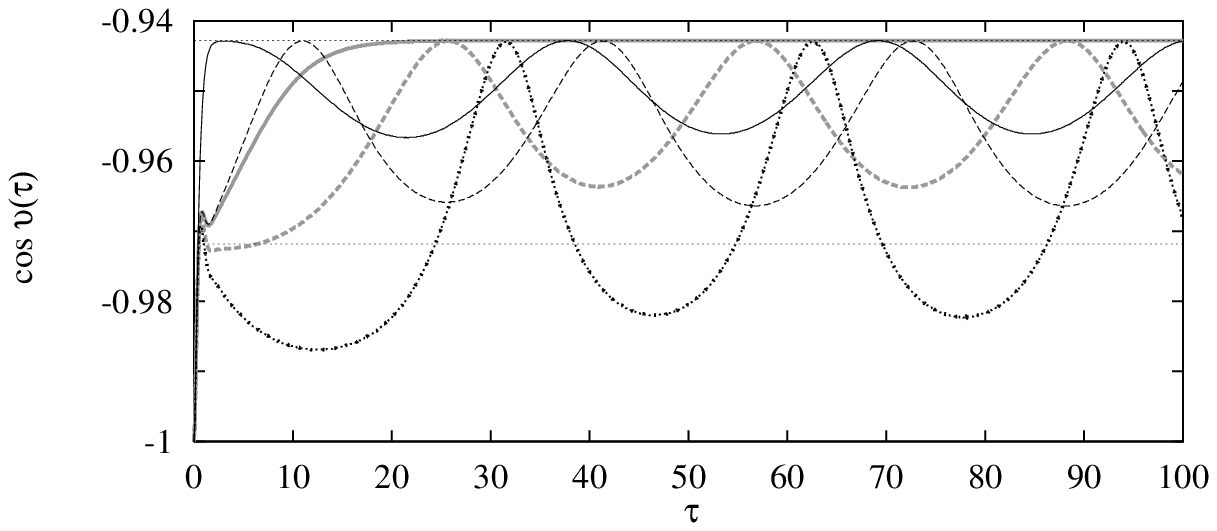,height=6cm,angle=0}}
\caption{
Time dependence of the total occupation (top) 
and of the relative site occupation difference (bottom)
for $g=3$, $p=3$ and $r=0.1$. In the bottom plot for the polaron (--) 
solution beside the relative site occupation 
the time dependence of the lower adiabatic state is displayed 
with sparse bold dots.
}
\label{3_3_0.1}
\end{figure}

\clearpage\vspace*{4cm}\begin{figure}
\centerline{\psfig{figure=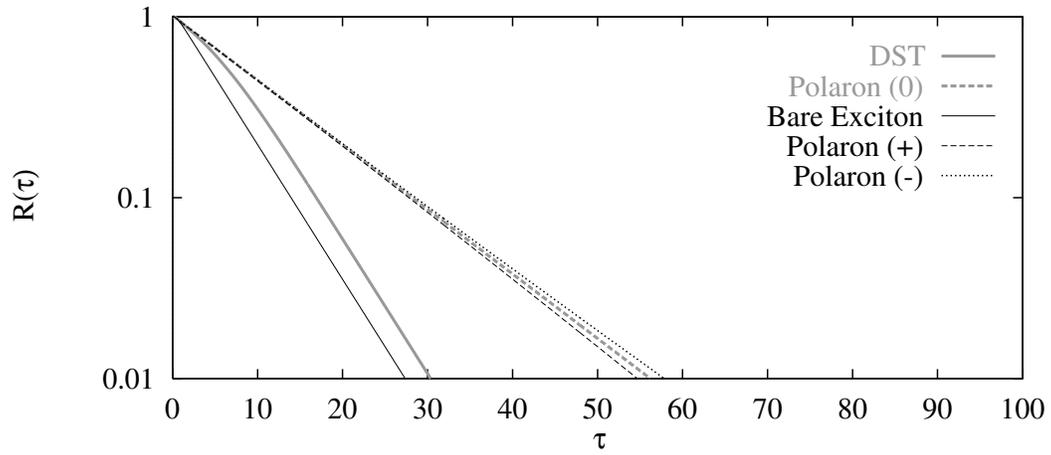,height=6cm,angle=0}}
\caption{
Time dependence of the total occupation 
for $g=3$, $p=3$ and $r=0.001$.
}
\label{3_3_0.001}
\end{figure}

%%%%%%%%%%%%%%%%%%%%%%%%%%%%%%%%%%%%%%%%%%%%%%%%%%%%%%%%%%%%%%%%%%%%%%

\end{document}